\documentclass[twocolumn,prd,amssymb,amsmath,preprintnumbers,floatfix,aps,nofootinbib,10pt]{revtex4-1}
\pdfoutput=1
\usepackage{dcolumn,graphicx,bm,psfrag}
\usepackage[colorlinks=true,citecolor=blue,urlcolor=black,linktocpage=true,
linkcolor=blue]{hyperref}
\usepackage{ dsfont }
\usepackage{multirow}
\usepackage{bm}
\usepackage{enumerate}

\newcommand{\gev}{\,\textrm{GeV}}
\newcommand{\tev}{\,\textrm{TeV}}
\newcommand{\mev}{\,\textrm{MeV}}
\newcommand{\kev}{\,\textrm{keV}}
\newcommand{\ev}{\,\textrm{eV}}

\newcommand{\hz}{\,\textrm{Hz}}
\newcommand{\mhz}{\,\textrm{MHz}}
\newcommand{\nm}{\,\textrm{nm}}
\newcommand{\gbl}{g_{B-L} }
\newcommand{\tn}[1]{\ensuremath{\textnormal{#1}}}

\newcommand{\mn}{m\nu^{AA'}}
\newcommand{\dr}{\delta\langle r^2 \rangle_{AA'}}
\newcommand{\mnvec}{\overrightarrow{m\nu}}
\newcommand{\mmuvec}{\overrightarrow{m\mu}}
\newcommand{\mrvec}{ \overrightarrow{m \delta \langle r^2 \rangle}}

\begin{document}

\title{\textbf{\Large Constraining New Physics Models with Isotope Shift Spectroscopy}} 

\author{\textbf{Claudia Frugiuele, Elina Fuchs, Gilad Perez and Matthias Schlaffer}}

\affiliation{Department of Particle Physics and Astrophysics, Weizmann Institute of Science, Rehovot, Israel}

\date{\normalsize  \small{May 9, 2017}}

\begin{abstract}
  Isotope shifts of transition frequencies in atoms constrain generic long- and intermediate-range
  interactions.  We focus on new physics scenarios that can be most strongly constrained by King
  Linearity Violation such as models with $B-L$ vector bosons, Higgs portal and chameleon.  With the
  anticipated precision, King Linearity Violation has the potential to set the strongest laboratory
  bounds on these models in some regions of parameter space. Furthermore, we show that this method
  can probe the couplings relevant for the protophobic interpretation of the recently reported Be
  anomaly. We extend the formalism to include an arbitrary number of transitions and isotope pairs
  and fit the new physics coupling to the currently available isotope shift measurements.
\end{abstract}

\maketitle
\section{Introduction}
\label{sec:intro}

The standard model of particle physics (SM) is one of the most successful scientific theories.  Yet,
it cannot be a complete description of nature because, for example, it does not provide a viable
dark matter candidate and cannot account for the observed baryon asymmetry of the universe.  Yet,
the energy scale associated with new physics (NP) is unknown and therefore the experimental program
for physics beyond the SM should be as broad as possible.  Colliders are one of the main tools to
study elementary particles and the LHC is pushing forward the energy frontier. A complementary and
vital role is played by low-energy, precision and/or high-intensity experiments, which require a
joint effort of the particle, atomic, and nuclear physics communities.  The MeV-GeV scale is already
efficiently probed by a variety of high-intensity experiments \cite{Alexander:2016aln, Essig:2009nc,
  Essig:2010xa, Batell:2009yf, Batell:2009di, Izaguirre:2014dua, Dobrescu:2014ita, Coloma:2015pih,
  Frugiuele:2017zvx,Flacke:2016szy}.  The existence of new sub-MeV degrees of freedom can instead be probed by both
astrophysical observations and precision experiments.  In this context, atomic physics observables
play an important role.  For example, atomic precision measurements can be used to constrain
interactions beyond the SM (BSM), see e.g.\ Refs.~\cite{Jaeckel:2010xx,Ficek:2016qwp}.  In a broader
context, atomic physics probes have been used to test the violation of fundamental laws such as
parity (see e.g.\ Refs.~\cite{Porsev:2009pr, Tsigutkin:2009zz, PhysRevA.81.032114, Leefer:2014tga,
  Versolato:2011zza, Dzuba:2012kx, Roberts:2014bka, Jungmann:2014kia}), Lorentz
symmetry~\cite{Dzuba:2015xva, Flambaum:2015rwt} and even the time variation of fundamental constants
of nature~\cite{Godun:2014naa, Rosenband1808}.\par
A new proposal to constrain NP using isotope shift measurements was presented in
Ref.~\cite{Delaunay:2016brc}. In Ref.~\cite{Berengut:2017zuo} this proposal was detailed and it was
shown that it can constrain new light degrees of freedom mediating long- and intermediate-range
spin-independent interactions between electrons and neutrons. These new interactions cause a
frequency shift that is factorized to a high degree into a product of a function that solely depends
on the electronic degrees of freedom and a function that depends on the nuclear physics ones.
Within the validity of this factorization, a linear relation between isotope shifts of different
transitions is obtained. This is known as \emph{King linearity}~\cite{King:63,King:13}. New
interactions mediated by light mediators that couple electrons to neutrons generally lead to a
non-linear relation~\cite{Delaunay:2016brc}. We shall denote such an effect as King Linearity
Violation (KLV).  The absence of a deviation from linearity allows to constrain the New Physics
contributions.\par
Existing measurements of isotope shifts cannot probe so far unconstrained regions of parameter
space, but the projected sensitivity allows to explore regions presently left unconstrained.  The
goal of this work is to investigate how this statement applies to specific models where couplings to
other SM particles (in addition to the electron and neutron couplings) and thereby additional
constraints become relevant.

The paper is organized as follows. In Sec.\,\ref{sec:NPIS} we introduce the notation and provide a
brief summary of Ref.~\cite{Berengut:2017zuo}.  In Sec.~\ref{sec:fit} we generalize the formalism
and provide a fit of NP interactions to the available data sets of Ca$^+$ and Yb.  As an
application, we identify models for which KLV constraints are relevant, such as a new gauge boson
$Z'_{B-L}$, the Higgs portal, chameleon models and protophobic models in the context of the recently
observed $^8$Be anomaly, and we discuss the implications before concluding in Sec.~\ref{sec:concl}.

\section{Probing new physics via isotope shift measurements}
\label{sec:NPIS}

Consider a narrow atomic transition $i$ between two atomic states and two even isotopes $A$ and
$A'$.  The isotope shift (IS) is defined as the difference of the transition frequencies,
$\nu_i^{AA'}\equiv \nu_i^{A}-\nu_i^{A'}$. The leading contributions to the IS stem from two sources:
the mass shift (MS) and field shift (FS). The former arises from the mass difference of the isotopes
$A$ and $A'$. It can be factorized into an electronic coefficient $K_i$, which only depends on the
transition $i$, and the isotope-dependent reduced mass given by
\begin{equation}
  \mu_{AA'}= \frac{1}{m_A}-\frac{1}{m_{A'}}\,,\label{mu}
\end{equation}
which is measured at high precision. The FS originates from the different volumes of the two
isotopes. At leading order it also factorizes into the electronic, isotope independent coefficient
$F_i$ and the charge radius variance,
$\delta\langle r^2 \rangle_{AA'} \equiv \langle r^2_A\rangle - \langle r^2_{A'} \rangle$, where
$r_A$ is the nuclear charge radius of isotope $A$. In contrast to $\mu_{AA'}$, $\dr$ is subject to
large experimental uncertainties.  The composition of the IS in terms of products of purely
electronic and purely nuclear quantities is referred to as factorization~\cite{King:63}.  As a
result, these two leading contributions amount to the total IS as
\begin{align}
  \label{eq:nuAApSM}
  \nu_i^{AA'} 
  =	K_i\, \mu_{AA'} + F_i\, \delta\langle r^2 \rangle_{AA'} +\ldots
  \, ,
\end{align}
where the first term represents the MS and the second one the FS~\cite{King:63,King:13}. The dots
denote possible higher order corrections and NP contributions which we will discuss below. It is
useful to normalize the frequency shifts by the reduced mass $\mu_{AA'}$ to obtain the so-called
\textit{modified} isotope shifts, $m \nu_i^{AA'} \equiv \nu_i^{AA'}/\mu_{AA'}$, which we will use in
the following. As a consequence, the mass shift is reduced to the electronic factor $K_i$ whereas
the FS factor $F_i$ is multiplied by the modified charge radius variance,
$m\dr\equiv \dr/\mu_{AA'}$.\par
When considering several pairs of isotopes, the modified Eq.~(\ref{eq:nuAApSM}) can be written in
vectorial form as
\begin{equation}
  \label{eq:nuAApVec}
  \mnvec_i = K_i\, \mmuvec + F_i\,\mrvec\,,
\end{equation}
where each line corresponds to one set of isotopes.  For the example of four isotopes combined to
three isotope pairs $\left\lbrace A, A_a\right\rbrace$, where $a=1,2,3$ and $A$ is the reference
isotope, the IS vector of transition $i$ is given by
$\mnvec_i = ( m \nu_i^{AA_1},m \nu_i^{AA_2}, m\nu_i^{AA_3})$, and $\mrvec$ accordingly.  The mass
shift vector is denoted by $\mmuvec = (1,1,1)$.\par

With measurements of two transitions $i=1,2$ the unknown charge radius distribution can be replaced
by measured quantities. Solving Eq.~(\ref{eq:nuAApVec}) with $i=1$ for $\mrvec$ and replacing it in
the equation with $i=2$ leads to
\begin{align}
  \label{eq:modKing}
  \mnvec_2  = K_{21}\,\mmuvec + F_{21}\, \mnvec_1\,,
\end{align}
with $F_{21}\equiv F_2/F_1$ and $K_{21}\equiv K_2-F_{21}K_1$.  Hence, this replacement gives rise to
a linear dependence between the two sets of modified frequency shifts $\mnvec_{1,2}$, referred to as
King linearity~\cite{King:63}.\par

In order to quantify the observed linearity, we define a measure of
nonlinearity~\cite{Berengut:2017zuo},
\begin{align}
  \label{eq:NL}
  {\rm NL} = ( \mnvec_1 \times  \mnvec_2 ) \cdot \mmuvec \,,
\end{align}
which corresponds to the volume of the parallelepiped spanned by the vectors $\mnvec_{1}$,
$\mnvec_{2}$ and $\mmuvec$ (for illustration see Ref.~\cite{Berengut:2017zuo}).  King linearity is
considered to hold if the measure NL is smaller than its uncertainty
$\delta {\textrm{NL}}$\footnote{At the present level of experimental accuracy, the uncertainties on
  the isotope masses are smaller by several orders of magnitude than those of the frequency shifts
  (e.g.~$\mathcal{O}(10^{-5})$ smaller for Yb masses~\cite{Myers2013107} with the present IS
  accuracy of $0.1-1\,$MHz~\cite{PhysRevA.94.052511,PhysRevA.20.239}). Therefore we will neglect
  them in our numerical evaluation. Once the uncertainties of IS measurements will be significantly
  reduced, the mass uncertainties will have to be taken into account.}. In several atoms and ions,
King linearity has indeed been established within the experimental uncertainty of $\sigma=0.1\mhz$
on the IS, see e.g.~Refs.~\cite{PhysRevLett.115.053003, Shi2016,PhysRevA.94.052511,
  PhysRevA.20.239, PhysRevA.59.3513}.

In Ref.~\cite{Berengut:2017zuo} it was shown that new physics contributions from light bosons
interacting with electrons and neutrons can lead to a deviation from the linear relation in
Eq.~(\ref{eq:modKing}). Thereby, the observation of linearity allows to set bounds on the mass and
coupling of a possible new force mediator.

To be specific, a new physics contribution is added as a third term to Eq.~(\ref{eq:nuAApSM})
\begin{align}
  \label{eq:nuAAp}
  \mnvec_i = K_i\, \mmuvec + F_i\, \mrvec  + y_ey_n X_i\, \vec h\,,
\end{align}
where $y_e$, $y_n$ are the couplings of a new boson to electrons and neutrons, respectively.
Furthermore, we have introduced the electronic NP factor $X_i$ and the reduced isotope dependence
$\vec h$. Both of them are model-dependent; a specific expression is given below. Proceeding as in
the SM case, one can express $\mnvec_2$ as a function of $\mnvec_1$, yielding
\begin{align}
  \label{eq:KingNL}
  \mnvec_2  = K_{21} \mmuvec+ F_{21}  \mnvec_1   + y_ey_n \vec h \left( X_2 - X_1 F_{21} \right)\,.
\end{align}
Thus, NP can break King linearity. For unit coupling, the NP contribution to NL is given by the
projection of $\vec h$ onto the normal vector of the King plane,
\begin{equation}
  \label{NLnp}
  { \rm NL_{NP}}=\left[ \mmuvec \times \left( X_2 \,\mnvec_1 - X_1 \,\mnvec_2\right)\right] \cdot \vec h\,.
\end{equation}
${ \rm NL_{NP}}$ vanishes if
\begin{enumerate}[(i)]
\item NP mediates a short-range interaction, shorter than the nuclear size. In this case the
  electronic parameter $X_i$ becomes proportional to the electronic parameter of the FS, namely
  $X_i\propto F_i$ so that the bracket in Eq.~(\ref{eq:KingNL}) vanishes or
\item the isotope-dependent NP contribution $ \vec h$ is proportional either to $ \mmuvec$ or to the
  reduced charge radius $\mrvec$, such that the NP contribution can be absorbed in a redefinition of $K_{21}$ or $F_{21}$, respectively.
\end{enumerate}\par
Finally, solving the set of equations in Eq.~(\ref{eq:KingNL}) determines the central value of
$y_ey_n$ needed to yield a particular data set \{$\mnvec_1$, $\mnvec_{2}$, $\mmuvec$\},
\begin{align}
  \label{eq:alpNP}
  y_e y_n &=  \frac{ \rm{NL}} { \rm{NL_{NP}} } \,.
\end{align}
The interval of $y_ey_n$ can be obtained via error propagation of the uncertainties on the involved
quantities.  In case of linearity, $y_ey_n$ is compatible with zero and the method reaches its
maximal sensitivity, whereas if nonlinearity is found a bound can be set with the experimental
resolution at which nonlinearity emerges.  In the following we will adopt the same approach as in
Ref.~\cite{Berengut:2017zuo} based on the best-case projection where linearity holds up to the
experimentally achievable precision.\par

Indeed, nonlinearity cannot only arise from NP, but also from SM higher-order contributions.  The
dominant effects are expected as corrections to the FS operator (see Refs.~\cite{PalmerStacey:1982,
  PhysRev.188.1916, SmExp:1981, 0022-3700-20-15-015, PhysRevA.31.2038} for relevant
discussions). However, these estimates are not tailored to the most promising elements and
transitions.  Thus, in order to fully exploit the KLV potential to probe NP interactions, a
significant improvement of the atomic theory input will be crucial to match the experimental
precision.\par

In the remainder of the paper, we will consider NP interactions that couple linearly to the SM
fermions. Hence the isotope-dependent NP part takes the form
\begin{equation}
  \label{eq:h}
  h_{AA'} = \frac{A-A'}{\mu_{AA'}} \simeq -AA'{\rm amu}\,,
\end{equation}
where in the last step we approximated $m_A \simeq A\, {\rm amu}$ and amu denotes the atomic mass
unit. Therefore, in this approximation, $\vec h$ can be written as
$\vec h = -A\, \vec A' \,\textnormal{amu}$ with $\vec A'=(A_1, A_2,A_3)$. We will further assume a
Yukawa-like potential of the NP interaction, mediated by a boson $\phi$ of mass $m_\phi$,
\begin{align}
  \label{eq:VNP}
  V_{{\rm NP}}(m_{\phi }, r) &= \frac{y_e y_n}{4\pi} (A-Z) \frac{e^{m_{\phi} r} }{r}\,.
\end{align}
In the massless limit, $m_{\phi}\ll (1+n_e)/a_0$ where $a_0$ is the Bohr radius and $n_e$ the
ionization number, the electronic NP factors $X_i$ can be expressed as~\cite{Berengut:2017zuo}
\begin{align}
  \label{eq:Ximassless}
  \left. X_i \right|_{m_\phi=0} 
  \approx 
  \frac{1}{2\pi \alpha}
  \left( \frac{E_{b} }{ Z_{\rm eff}^b} - \frac{E_a }{ Z_{\rm eff}^a} \right)\,,
\end{align}
with $\alpha$ being the fine structure constant.  For the transition $i = a\rightarrow b$ between
the energy levels $E_a$ and $E_b$, the effective nuclear charges $Z^{\psi}_\textnormal{eff}$ account
for the partial shielding of the nuclear charge for a valence electron at the states $\psi =a, b$,
respectively. For later use, we define
\begin{align}
 x_i &= X_i\, A\, {\rm amu} \label{eq:xi}\,,\\
 x_{ij} &= x_i -F_{ij}\, x_j \label{eq:xji}\,.
\end{align}
In the numerical evaluation in Sec.~\ref{sec:fit} we will use this $Z_{\rm eff}$ approximation, see
Ref.~\cite{Berengut:2017zuo} for more details.

\section{Fit of the NP coupling}
\label{sec:fit}

The formalism proposed in Ref.~\cite{Berengut:2017zuo} and summarized in Sec.~\ref{sec:NPIS} is
constructed for the minimal case of two transitions and three isotope pairs.  However, at present
there are already more measurements at comparable accuracy available in some systems, such as three
transitions in Ca$^+$~\cite{PhysRevLett.115.053003,Shi2016} and four independent isotope pairs in Yb
~\cite{PhysRevA.94.052511,PhysRevA.20.239}. Hence the system is overconstrained and a fit of the NP
coupling is necessary. In the following we will therefore perform a $\chi^2$ fit to the data to
obtain a limit on the NP coupling $y_ey_n$.

\renewcommand{\arraystretch}{1.4}
\begin{table}[tbp]
  \begin{center}
    \begin{tabular}{|c|c l|c|c|c|c|}
      \hline
      	element	&&transition				&$\lambda\,[{\rm nm}]$	&$\sigma$\,[MHz] &\,$n$\,	&Ref.	\\\hline\hline
      \multirow{3}{*}{Ca$^+$}	     		&&$4S_{1/2} \rightarrow 4P_{1/2}$ (D1)~	&397		&0.1&3&\cite{PhysRevLett.115.053003}\\
      		&&$3D_{3/2} \rightarrow 4P_{1/2}$ 	&866		&0.1&3&\cite{PhysRevLett.115.053003}\\
      		&&$4S_{1/2} \rightarrow 4P_{3/2}$ (D2)~	&393		&0.1&3&\cite{Shi2016}\\
      \hline\hline
      \multirow{4}{*}{Yb}&& $6^1S_0 \rightarrow 6^1P_1$ 	&399&0.5	&4&\cite{PhysRevA.94.052511}\\
      && $6^1S_0 \rightarrow6^3P_1$ 	&555.65&0.5	&4&\cite{PhysRevA.20.239}\\
      &&$6^3D_2\rightarrow 6^1S_0$ &404	&10&3&\cite{PhysRevA.59.3513}\\
      &&$6^3D_1\rightarrow 6^1S_0$ &408	&2&3&\cite{PhysRevA.59.3513}\\
      \hline\hline
    \end{tabular}
    \caption{Measured transitions in Ca$^+$ and neutral Yb.  $\lambda$ denotes the wavelength of the
      transition in the reference isotope $A$, $\sigma$ the experimental uncertainty on the isotope
      shifts, and $n$ the number of available isotope pairs. In Ca$^+$, $A=40$ is the reference
      isotope and $A'=42,44,48$. In Yb, $A=174$ and $A'=(168,)\, 170, 172, 176$ for $n=3 (4)$. The
      Yb transitions with $\lambda=404\nm$ and $408\nm$ are omitted in the fit due to their lower
      current resolution.}
    \label{tab:transitions}
  \end{center}
\end{table}

\begin{table*}[tp] 
  \centering
  \begin{tabular}{|c|c|lll|c|}
    \hline
    ~element~ & ~omitting~ &~ $\left.y_e y_n\right|_\tn{min}\qquad\quad$ &~ $\left.y_e y_n\right|_\tn{best}\qquad\quad$ &~
                                                                                              $\left.y_e y_n\right|_\tn{max}$ & ~$\chi^2_\tn{min}$~\\\hline \hline
    \multirow{4}{*}{Ca$^+$}
     & 397\,nm                 & $-2.8\times 10^{-9}$ & $-6.7\times 10^{-10}$ & $+1.3\times 10^{-9}$ & 0.00\\
     & 866\,nm & $-9.2$              & $+1.0$                & $+8.8$              & 0.00\\
     & 393\,nm                 & $-2.8\times 10^{-9}$ & $-6.1\times 10^{-10}$ & $+1.5\times 10^{-9}$ & 0.00\\ \cline{2-6}
     & --                                       & $-2.8\times 10^{-9}$  & $-6.4\times 10^{-10}$ & $+1.3\times 10^{-9}$ &0.04\\\hline \hline
    \multirow{5}{*}{Yb} & 168  & $-2.8\times 10^{-9}$ & $-9.3\times 10^{-10}$ & $+9.9\times 10^{-10}$ & 0.00\\
                        & 170  & $-2.6\times 10^{-9}$ & $-7.6\times 10^{-10}$ & $+1.1\times 10^{-9}$ & 0.00 \\
                        & 172  & $-2.8\times 10^{-8}$ & $+1.2\times 10^{-8}$ & $+6.5\times 10^{-8}$ & 0.00 \\
                        & 176  & $-1.9\times 10^{-8}$ & $-5.1\times 10^{-9}$ & $+9.7\times 10^{-9}$ & 0.00 \\ \cline{2-6}
                        &  --  & $-2.6\times 10^{-9}$ & $-8.1\times 10^{-10}$ & $+1.1\times 10^{-9}$ & 0.34 \\\hline \hline
  \end{tabular}
  \caption{Minimal value of $\chi^2$ as well as upper and lower 95\% CL bounds and the best fit
    value of the product of the couplings as determined by the fit for $m_\phi=0$.}
  \label{tab:fit_limits}
\end{table*}

Under the assumptions made in Eq.~(\ref{eq:h}) and Eq.~(\ref{eq:Ximassless}), Eq.~(\ref{eq:KingNL})
can be written for any two transitions $i$ and $j$ as
\begin{equation}
  \label{eq:NP_king_relation}
  \mnvec_i = K_{ij} \mmuvec + F_{ij}\mnvec_j + y_e y_n\, x_{ij}\, \vec A'\,,
\end{equation}
where $x_{ij}$ is given in Eq.~(\ref{eq:xji}).  This equation defines a family of parallel lines
whose intercept depends on the new physics couplings $y_e y_n$ and the isotope pair via the third
term. The lines live in the isotope shift space where each dimension corresponds to an isotope
pair. When combining $m$ transitions, we obtain for each isotope pair $\{A,A'\}$ a line in the
$m$-dimensional space
\begin{equation}
  \label{eq:Fit_equation}
  \begin{pmatrix}
    \mn_i\\
    \mn_j\\
    \vdots\\
    \mn_m
  \end{pmatrix} =
  \begin{pmatrix}
    0\\
    K_{ji}\\
    \vdots\\
    K_{mi}
  \end{pmatrix}
  + \mn_i
  \begin{pmatrix}
    1\\
    F_{ji}\\
    \vdots\\
    F_{mi}
  \end{pmatrix}
  + y_e y_n A'
  \begin{pmatrix}
    0\\
    x_{ji}\\
    \vdots\\
    x_{mi}
  \end{pmatrix}
\end{equation}
that can be fitted to the measured isotope shifts. For later convenience we write this equation as
\begin{equation}
  \label{eq:fitvector}
  \vec P^{AA'} = \vec K + \mn_i \vec F + y_e y_n A' \vec x\,,
\end{equation}
where the components of the above vectors follow from Eq.~(\ref{eq:Fit_equation}).\par
For the fit we construct a $\chi^2$ function and marginalize over the entries of the vectors  $\vec K $ and
$\vec F $. Since the measured isotope shifts exhibit similar uncertainties in all transitions, a
multi-dimensional $\chi^2$ that includes the uncertainties of all transitions needs to be
constructed (see e.g.~Ref.~\cite{Hogg:2010yz}). For simplicity we assume that the uncertainties of the
measurement are not correlated. In this case the contribution $\chi^2_{AA'}$ of the pair $\{A,A'\}$
to the $\chi^2$ function is given by
\begin{equation}
  \label{eq:local_chi2}
  \chi^2_{AA'}=\sum_i \left(\frac{d^{AA'}_{i}}{\sigma\mn_i}\right)^2\,,
\end{equation}
where the sum runs over all transitions.  Here, $\sigma\mn_i$ is the uncertainty of the respective
IS measurement, and $d^{AA'}_{i}$ is the $i$-th component of the vector connecting the measured
point $\vec P^{AA'}=(\mn_i,\mn_j,\ldots,\mn_m)$ to the line defined by Eq.~(\ref{eq:Fit_equation}),
\begin{equation}
  \label{eq:dvector}
  \vec d^{AA'}= \left(\vec P^{AA'} - \vec c_0^{AA'}\right) - \hat n\cdot \left[\hat n \cdot \left( \vec P^{AA'}-\vec c_0^{AA'}\right)\right]\,,
\end{equation}
with $\vec c_0^{AA'}=\vec K + y_e y_n A' \vec x$ and $\hat n=\vec F/|\vec F|$. The full $\chi^2$
function is given by summing over all isotope pairs.\par
By construction, the IS of the transition that appears in the right-hand side of
Eq.~(\ref{eq:Fit_equation}) seemingly has a special role. This is, however, not the case as the
distance between a point and a line in an $m$-dimensional space is invariant under the permutation
of coordinates. When minimizing the $\chi^2$ we have checked the consistency of our computation and its numerical stability by
verifying that all permutations of the isotope shifts yield comparable results. We obtain the 95\%
confidence level limits on $y_ey_n$ shown in Table~\ref{tab:fit_limits}. The near degeneracy between
two of the transitions in Ca$^+$ is reflected in the extremely weak limit in the case of omitting
the non-degenerate transition of $\lambda=866\nm$. Moreover this explains why the limit hardly
becomes more stringent when including all three transitions. For Yb the limits get in general weaker
by a factor $\mathcal{O}(1)$ to $\mathcal{O}(10)$ when one isotope shift measurement is dropped. The
removal of $A'=172$ leads to the weakest bound.\par

Omitting one transition of Ca$^+$ or one isotope pair of Yb leads to the minimal case of $m=2$
transitions and $n=3$ isotope pairs where Eq.~(\ref{eq:NP_king_relation}) is exactly
solvable. 
Hence, the theory parameters $F_{21}$, $K_{21}$ and $y_ey_n$ can be chosen such that the theory predictions of the modified isotope shifts reproduce exactly the measured ones.
This is reflected by the vanishing $\chi^2_{\rm min}$ in Tab.~\ref{tab:fit_limits}.

\section{Implications for BSM models}
\label{sec:bsm-models}

In Ref.~\cite{Berengut:2017zuo} the sensitivity of KLV was compared to other measurements in a
model-independent way. In the following we will translate the KLV bounds into bounds on the
parameters of various BSM models and compare them to existing constraints. In addition we will
explore the sensitivity of the near-future KLV projections with Ca$^+$ D-states, Sr$^+$, Sr/Sr$^+$,
and Yb$^+$ as reported in Ref.~\cite{Berengut:2017zuo}. We
focus on those models that can be best probed by KLV measurements.
While we discuss in detail the various constraints on the $B-L$ model in the full mass range relevant for KLV, we highlight promising mass values in Higgs portal and chameleon models. Most bounds on the $B-L$ model can be translated also to these models by rescaling.
Furthermore, we present updated bounds on the protophobic interpretation of the Be anomaly.

\subsection{\texorpdfstring{\textbf{\emph{Z\,$'$}} vector boson from
    \textbf{\emph{U(1)}}$_\textnormal{\textbf{\emph{B-L}}}$} {Z' vector boson from U(1)\_(B-L)}}
\label{sec:B-L}

One of the frequently studied abelian extensions of the SM gauge group is gauging the difference of
baryon and lepton number, $B-L$. Under this additional interaction all quarks therefore have the
same charge $z_q=1/3$ and all leptons $z_l=-1$. The group is made anomaly-free by introducing a
right-handed neutrino for each family.  In this model, the coupling $g_{B-L}$ of the new vector
boson $Z'$ is purely vectorial and of equal strength for electrons and neutrons, hence KLV is a
promising method to probe this kind of NP interaction.

\begin{figure}[tp]
  \begin{center}
    \includegraphics[width=\columnwidth]{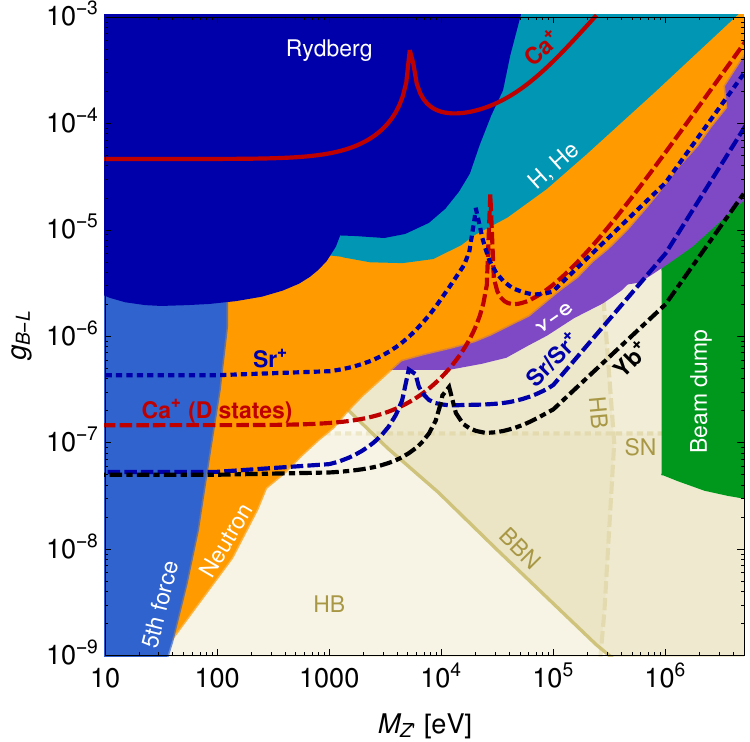}
  \end{center}
  \caption{Constraints on a $Z'$ gauge boson from $U(1)_{B-L}$.  KLV bound from existing IS data:
    Ca$^+$ with uncertainty $\sigma\simeq 0.1\mhz$ (397\,nm
    vs. 866\,nm~\cite{PhysRevLett.115.053003}, solid red line).  KLV projections for $\sigma=1\hz$
    assuming linearity in Ca$^+$ ($S\to D$ transitions, red, dashed), Sr$^+$ (blue, dotted),
    Sr/Sr$^+$ (blue, dashed), and Yb$^+$ (black, dash-dotted)~\cite{Berengut:2017zuo}.  For
    comparison, bounds from fifth-force searches via the Casimir
    effect~\cite{Bordag:2001qi,bordag2009advances} (blue), neutron
    scattering~\cite{Leeb:1992qf,Nesvizhevsky:2007by,Pokotilovski:2006up} (orange), Rydberg
    states~\cite{PhysRevLett.61.2285,Karshenboim:2010cg,Karshenboim:2010ck} (dark blue), energy
    level shifts in H and He~\cite{Jaeckel:2010xx} (turquoise), $\nu-e$ scattering at GEMMA and
    Borexino~\cite{Harnik:2012ni} (purple), and beam dump experiments~\cite{Essig:2009nc,
      Essig:2010xa, Andreas:2012mt} (green).  Astrophysical and cosmological probes (beige):
    supernova 1987A with $\mathcal{O}(1)$ uncertainties~\cite{Yao:2006px,
      Raffelt:2012sp,Blum:2016afe} (SN, the area below the dotted line), horizontal branch
    stars~\cite{Yao:2006px, Grifols:1986fc, Grifols:1988fv, Hardy:2016kme, Redondo:2013lna} (HB, the
    area left of the dashed line) and BBN via $N_{\rm eff}$~\cite{Ahlgren:2013wba, Heeck:2014zfa}
    (the area above the solid line).  }
  \label{fig:ZBL}
\end{figure}

In Fig.~\ref{fig:ZBL} we compare the KLV bounds and projections from different atoms and ions to
existing constraints in the mass range of $M_{Z'}\sim$ 10\,eV to a few MeV. For other overviews
collecting bounds on this model and related models see e.g.~Refs.~\cite{Jaeckel:2010ni,
  Jaeckel:2010xx, Harnik:2012ni, Heeck:2014zfa}.

\subsubsection{Laboratory bounds}
\label{sec:laboratory-bounds}

The existence of a fifth force is severely constrained for a mass $M_{Z'}\lesssim 100\ev$ by
experiments testing the Casimir effect~\cite{Bordag:2001qi, bordag2009advances}.

In contrast to KLV,
other atomic precision measurements 
such as energy level shifts in Rydberg
states~\cite{PhysRevLett.61.2285,PhysRevLett.61.2285,Karshenboim:2010cg,Karshenboim:2010ck} and in
$s$- and $p$-states of atomic H and hydrogen-like He$^+$~\cite{Jaeckel:2010xx}
provide bounds on $ y_p y_e $ where $ y_p$ is the proton coupling.
In the massless limit, $M_{Z'}\ll (1+n_e)/a_0$, the NP potential probed by these observables simplifies to a Coulomb potential.
In this case the NP interaction is absorbed by a redefinition of the fine-structure constant
$\alpha$, resulting in a weakening of the bounds. 
Due to its sensitivity to $y_ey_n$, KLV is not affected by this redefinition so that its bound remains constant in the massless limit
and is the strongest among the atomic spectroscopy bounds for mediator masses below $0.3\ev$.
The intersection of the Ca$^+$ and Rydberg bound was determined following Ref.~\cite{Karshenboim:2010ck} and lies below the mass range shown in Fig.~\ref{fig:ZBL}. 
Yet, one needs to keep in
mind that for $M_{Z'}\leq 0.3\ev$ also other constraints apply, 
such as from the Casimir effect mentioned above or from tests for a deviation from the Coulomb force, see e.g.\
Ref.~\cite{Jaeckel:2010ni}. 

\begin{figure}[tbp]
  \centering
  \includegraphics[width=\columnwidth]{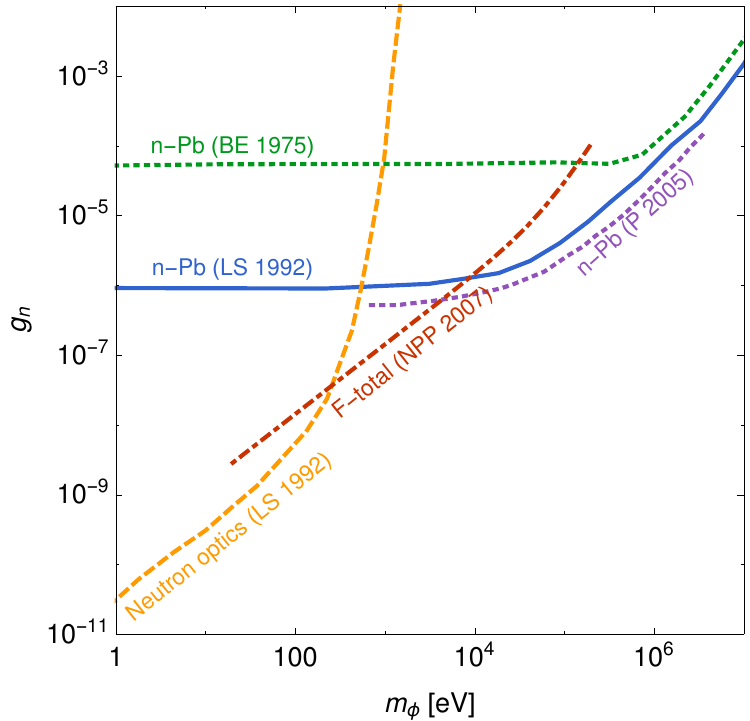}
  \caption{Existing bounds on the neutron coupling $g_n$ of a new boson $\phi$~ from the
    neutron-electron scattering length in Pb, Bi and noble gases denoted as neutron
    optics~\cite{Leeb:1992qf} (orange, dashed); $n$-$^{208}$Pb scattering at neutron energies of
    $E_n\sim 1\kev$-$26\kev$~\cite{Barbieri:1975xy} (green, dotted),
    $10\ev$-$10\kev$~\cite{Leeb:1992qf} (blue, solid) and up to $20\kev$ including interference of
    resonant and non-resonant amplitudes~\cite{Pokotilovski:2006up} (purple, dotted); and the
    comparison of the total to the forward scattering cross section of neutrons on
    nuclei~\cite{Nesvizhevsky:2007by} (red, dash-dotted). For discussion see
    Sect.~\ref{sec:laboratory-bounds}.}
  \label{fig:neutron_bounds}
\end{figure}

Neutron scattering is a powerful probe of the interaction between new bosons and neutrons over a
wide mass range.  Among the neutron scattering experiments, neutron optics~\cite{Leeb:1992qf}
provides the strongest constraint on $g_n$, in this model equivalent to $\gbl$, in the mass range of
$M_{Z'}\lesssim 500\ev$. For $500\ev\lesssim M_{Z'}\lesssim 5\kev$, the comparison of the total to the forward
scattering cross section~\cite{Nesvizhevsky:2007by} is most sensitive.  Above $M_{Z'}\sim 5\kev$,
the neutron-lead ($n$-Pb) scattering~\cite{Pokotilovski:2006up} sets the strongest bound. This
method is based on the proposal by Ref.~\cite{Barbieri:1975xy} whose bounds are superseded by the
ones reported in Refs.~\cite{Pokotilovski:2006up,Leeb:1992qf}. The collection of the various bounds
is shown in Fig.~\ref{fig:neutron_bounds}, the limit presented in Fig.~\ref{fig:ZBL} shows the best
bound for each mass.
When comparing to KLV, the considerable uncertainties on the neutron scattering bounds need to be
kept in mind~\cite{Wissmann:1998ta, Nesvizhevsky:2007by, Antoniadis:2011zza, TuckerSmith:2010ra}. In
particular, the uncertainties related to the electron-neutron scattering length, various nuclear
inputs, and the missing higher-order terms in the neutron-scattering cross section, are not easy to
determine precisely. Similarly, the bounds inferred for masses far higher than the neutron energies
of $E_n\sim10\kev$~\cite{Leeb:1992qf} are also subject to large uncertainties. In addition, in the
derivation of the various neutron bounds it is assumed that the NP contribution to the
neutron-electron interaction is either absorbed in the corresponding measurement of the
neutron-electron scattering length, $b_{ne}$, or negligible~\cite{Nesvizhevsky:2007by}.

For masses above a few keV, the neutron bounds are exceeded by constraints that arise from
measurements of neutrino-electron ($\nu$-$e$) scattering. Measurements from
Borexino~\cite{Bellini:2011rx} and GEMMA~\cite{Beda:2009kx} have been interpreted in the $B-L$ model
in Ref.~\cite{Harnik:2012ni} and are the strongest laboratory bounds between $4\kev$ and $1\mev$.

The limit on $g_{B-L}$ via $g_e$ from the bound on the anomalous magnetic moment of the electron,
$(g-2)_e$, is less constraining than the neutron scattering bounds in the same mass
range~\cite{PhysRevLett.100.120801}; thus we omit it in Fig.~\ref{fig:ZBL}.

Above the electron threshold, i.e.\ $M_{Z'}\gtrsim1\mev$, and up to $M_{Z'}\lesssim 100\mev$,
high-intensity electron and proton beam dump experiments~\cite{Essig:2009nc, Essig:2010xa,
  Andreas:2012mt} provide the strongest bound on $\gbl$. For a review see
Ref.~\cite{Alexander:2016aln} and references therein.

\subsubsection{Astrophysical and cosmological bounds}
\label{sec:astrophysical-bounds}

A large part of the parameter space of a new light boson interacting with SM fermions is constrained
by astrophysical probes. In particular, bounds on the cooling rate in horizontal branch stars limit
the coupling of $Z'$ as long as its mass is within the thermal reach
($M_{Z'}\lesssim 350\kev$)~\cite{Yao:2006px, Grifols:1986fc, Grifols:1988fv, Hardy:2016kme, Redondo:2013lna}.  Here
we omit the corresponding bound from sun cooling since its excluded region is also covered by the
horizontal branch stars.  Furthermore, the energy loss in the core of the supernova SN 1987A
constrains the neutron coupling for masses of $M_{Z'}\lesssim 100\mev$~\cite{Yao:2006px,
  Raffelt:2012sp}, though with large uncertainties~\cite{Blum:2016afe}.

The coupling $g_{B-L}$ is also severely constrained by the effective number of neutrinos
$N_{\textrm{eff}}$~\cite{Ahlgren:2013wba, Heeck:2014zfa}. If a light mediator couples to neutrinos,
it can thermalize via the inverse decay $ \nu+ \bar \nu \rightarrow Z'$ or
$ \nu+ \bar \nu \rightarrow Z' Z'$, thus contributing to the energy density of the Universe. The
first process dominates for $1\ev \lesssim M_{Z'} \lesssim 1\mev$, and the second one for
$ M_{Z'} \ll 1\ev$.  Under the requirement that the NP contribution to $N_{\textrm{eff}}$,
$ \Delta_{\nu}$, fulfills $\Delta_\nu <1 $ at $T=1\mev$, a large parameter region is excluded.

Although the cosmological and astrophysical bounds exceed all KLV projections, a complementary
laboratory probe of this region will nevertheless be valuable, in particular because isotope shifts
are a very clean observable and the derived bounds are based on less model assumptions. E.g.\ the
constraint from $\Delta_\nu$ can be avoided by charging only the right-handed electrons under the
new gauge group. However, in such a model other strong constraints arise, for instance, from the
cancellation of a gauge anomaly in the case of an anomalous gauge group~\cite{Dobrescu:2014fca}. For
examples of UV-complete models of axial couplings with such phenomenology see
Refs.~\cite{Kozaczuk:2016nma, Kahn:2016vjr}.  Moreover, in this class of models the axial coupling
$g_A^e \neq 0.$ In presence of an axial coupling to electrons and vectorial couplings to quarks,
atomic parity violation (APV)~\cite{Bouchiat:2004sp} is a more sensitive probe than KLV.  For
instance for gauge bosons with $m_{Z'} \sim 0.1\mev$, the limit from APV \cite{Bouchiat:2004sp} is
several orders of magnitude stronger than the Yb$^+$ projection.  The same conclusion is obtained
for constraints on the mixing of a new light gauge boson $Z'$ with the SM
$Z$~\cite{Davoudiasl:2012ag}.

\subsubsection{KLV bounds and projections}
\label{sec:klv-bounds}

Due to the simple coupling structure of the $Z'$ boson, the limits from KLV are straightforwardly
obtained from Ref.~\cite{Berengut:2017zuo} by identifying $g_{B-L}^2=y_ey_n$. The existing bound
from Ca$^+$ is shown by the solid red line, the projections are shown by dashed lines.  While the
existing bound does not compete with other laboratory bounds, the projected bounds will extend the
reach of laboratory bounds in the mass range of $300\ev \lesssim M_{Z'} \lesssim 1\mev$, in
particular using Sr/Sr$^+$ and Yb$^+$.  Hence, KLV has the potential put the regions that are
currently only probed by astrophysical and cosmological observables under scrutiny.

\subsection{Higgs portal}
\label{sec:higgs-portal}

Another example for a new light mediator which gives rise to a spin-independent interaction between
electrons and neutrons is a scalar singlet mixed with the Higgs
boson~\cite{Patt:2006fw,OConnell:2006rsp}. This model can be linked to the solution of the hierarchy
problem via the relaxion mechanism, where the relaxion takes the role of the new scalar that mixes
with the Higgs boson~\cite{Flacke:2016szy,Choi:2016luu}.\par

We evaluate the KLV sensitivity to this class of models for a very light scalar $ \phi$ with mass
$ m_{\phi} \lesssim 5\kev$. In this case the KLV bound on the mixing angle $\theta_{h\phi}$ between
the singlet and the Higgs boson is given by
\begin{align}
  \label{eq:portal}
  \sin^2{\theta_{h\phi}} & \lesssim 2\cdot 10^{-6}\,\cdot\,\left[ \frac{4\cdot 10^{-9} }{y_ey_n}\right]\, \frac{\sigma}{\textrm{Hz}}\,,
\end{align}
where we assumed the strongest KLV projections for this mass given in Ref.~\cite{Berengut:2017zuo},
i.e.~Yb$^+$ with a precision $\sigma$ at the $1\hz$ level.  The quark and gluon contribution to the
neutron Yukawa coupling can be obtained using Refs.~\cite{Belanger:2008sj, Belanger:2013oya,
  Shifman:1978zn}.  For SM Yukawa couplings, the limit from neutron scattering is stronger by one
order of magnitude since the KLV observable depends on the electron Yukawa coupling that is much
smaller than the neutron Yukawa coupling. Yet, for models where the electron (neutron) coupling is
enhanced (suppressed) by at least one order of magnitude with respect to its SM value, KLV could set
a stronger bound than neutron scattering.  For assumptions regarding the bound from neutron
scattering see Sect.~\ref{sec:laboratory-bounds}.  Under these assumptions and if the electron
(neutron) coupling is enhanced (suppressed) by a factor of $\sim$10, KLV sets a stronger bound than
neutron scattering for $m_\phi\gtrsim 30\kev$ and the strongest bound of all in the region
$350\kev\lesssim m_\phi\lesssim 1\mev$. Below this range a stronger bound arises from horizontal
branch stars~\cite{Yao:2006px, Grifols:1986fc, Grifols:1988fv, Hardy:2016kme, Redondo:2013lna} and
above 1\,MeV beam dump experiments are more constraining.\par

A suppressed $y_n$ can arise for example in less minimal Higgs portal models, such as the leptonic
Higgs portal~\cite{Batell:2016ove}. In this model, the singlet $\phi$ couples to leptons with a
similar strength as the $125\gev$ Higgs boson, i.e.~$y_l^{\phi} \sim \frac{m_l}{v}$, while its
coupling to the quarks is suppressed compared to normal Higgs portal scenarios, hence neutron
scattering experiments loose sensitivity.

\subsection{Chameleon models}
\label{sec:chameleon}

The chameleon is a scalar field $\phi$ with an effective potential that depends on the density
$\rho$ of the environment~\cite{Khoury:2003aq,Khoury:2003rn}
\begin{equation}
  \label{eq:chameleon_potential}
  V_\tn{eff}=V(\phi)+\frac{\phi\rho}{M}\,,
\end{equation}
with $M$ being a mass scale characterizing the coupling of the chameleon to matter, and $V(\phi)$ is
chosen such that the mass of $\phi$ increases with increasing $\rho$. As a result the mass of the
scalar is heavy in a dense environment and light otherwise, which leads to a screening effect in
test masses. Therefore it can mediate a long-range force on cosmological scales but avoid
constraints from fifth-force experiments.\par

On atomic scales the chameleon can alter the energy levels of the electrons.  The relevant part of
the NP perturbation of the Hamiltonian that can be probed by KLV is given
by~\cite{Brax:2010jk,Brax:2010gp}
\begin{equation}
  \label{eq:chameleon_dH}
  \delta H \big |_{n}=-\frac{m_e m_N}{4\pi r M^2}\,,
\end{equation}
where $m_{e}$ is the electron mass, $m_N \simeq (A-Z) m_n$ the contribution of the neutrons to the
nucleus mass, and $r$ the distance to the nucleus.  In this expression we omit a possible screening
of the nucleus as it depends not only on the parameter space of the model but also on the
experimental setup \cite{Burrage:2014oza}.\par

Assuming the massless case, Eq.~(\ref{eq:chameleon_dH}) can be matched to Eq.~(\ref{eq:VNP}) and the
bounds on $y_e y_n$ can be easily translated into bounds on $M$
\begin{equation}
  \label{eq:chameleon_M}
  M>\sqrt{\frac{m_e m_n}{y_e y_n\big|_\tn{min}}}\approx 500\tev\approx 2.5\cdot 10^{-13}\,M_\tn{Pl}\,,
\end{equation}
where $M_\tn{Pl}$ is the reduced Planck mass and we used the Yb$^+$ KLV projection. This is stronger
by more than an order of magnitude than the current best bound from measurements of energy levels in
hydrogen and helium atoms \cite{Jaeckel:2010xx,Schwob:1999zz,Simon:1980hu}, which leads to a bound
of $M\gtrsim10\,\tev$ \cite{Burrage:2016bwy}. Depending on the parameter space, this can even be the
strongest bound on $M$. The current bound of $10\tev$ is stronger than the present KLV bound from
Ca$^+$ and Yb.

\subsection{Beryllium anomaly}
\label{sec:Beryllium}

Recently, a $6.8\,\sigma $ anomaly was reported in rare nuclear decays of
$\null^8$Be~\cite{Krasznahorkay:2015iga}. The anomaly arises in the iso-scalar transition
$\null^8\rm{Be}^* (1^+) \to \null^8\rm{Be}(0^+) + e^+ e^-$ as a bump in the distribution of the
opening angle of the emitted electron-positron pairs. This observation can be explained by the
emission of a particle $X$ with mass $m_X\approx 17\mev$ in the process
$ \null^8\textrm{Be}^* (1^+) \to \null^8\textrm{Be}(0^+) + X$, which subsequently decays into an
electron positron pair. The best agreement with observations is obtained for $X$ being a vector with
either axial or vectorial couplings to quarks and electrons \cite{Feng:2016jff, Feng:2016ysn,
  Kozaczuk:2016nma, Kahn:2016vjr}. It was noted in Refs.~\cite{Feng:2016jff, Feng:2016ysn} that the
vector-like interpretation necessitates protophobic couplings to quarks, or else it would be in
conflict with other observables. Therefore KLV can provide the necessary method to confirm or reject
this hypothesis.\par

\begin{figure}[tp]
 \begin{center}
  \includegraphics[width=0.95\columnwidth]{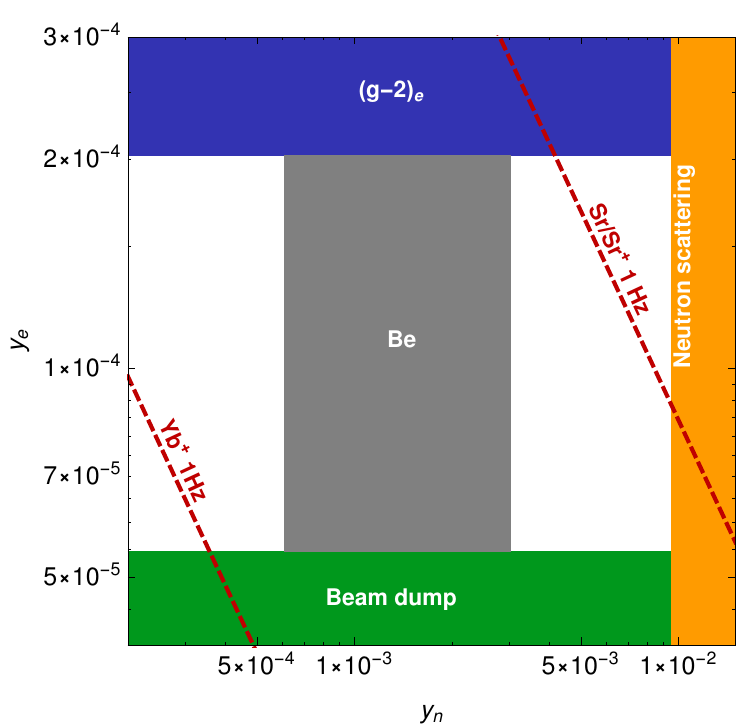}
  \caption{95\% CL bounds on $y_e$ and $y_n$ for a protophobic vector boson of mass
    $m_{X}=17\mev$. The gray region represents the required and allowed couplings to explain the
    $\null^8$Be anomaly. The dashed lines show the projected upper bounds on the couplings from KLV
    measurements in Sr/Sr$^+$ and Yb$^+$. }
  \label{fig:Beryllium_plot}
 \end{center}
\end{figure}

In Fig.~\ref{fig:Beryllium_plot} we present KLV projections and compare them to existing bounds on
$y_e$ and $y_n$ for a fixed mass of $m_{X}=17\mev$. The gray shaded area corresponds to the range of
the couplings that explains the observed excess and is not in conflict with other measurements.\par
The upper bound on $y_e$ comes from $(g-2)_e$ measurements. In contrast to Refs.~\cite{Feng:2016jff,Feng:2016ysn} the plotted bound represents the 95\% CL instead of the $3\,\sigma$ limit. 
Another upper bound is provided by KLOE-2~\cite{Anastasi:2015qla}. However even this 90\% CL bound is weaker
than the 95\% CL bound from the magnetic moment and therefore not shown. The lower bound on $y_e$
stems from beam dump experiments requiring that the new particle decays before it leaves the
detector. The strongest bound for $m_X=17\mev$ is provided by the E141
experiment~\cite{Riordan:1987aw} and interpreted in Ref.~\cite{Andreas:2012mt}. The latter corrected
Ref.~\cite{Essig:2009nc} that was used in Refs.~\cite{Feng:2016jff, Feng:2016ysn}.\par

An upper bound on $y_n$ is set by neutron-Pb scattering. The strongest constraint for $m_X=17\mev$
is provided by Ref.~\cite{Leeb:1992qf} which is stronger than the older derived in
Ref.~\cite{Barbieri:1975xy} and that was used in Refs.~\cite{Feng:2016jff,Feng:2016ysn}. The bound
shown in Fig.~\ref{fig:Beryllium_plot} is weaker by a factor of $\sqrt{A/(A-Z)}$ than the one
presented in Fig.~\ref{fig:neutron_bounds} due to the protophobic nature of the coupling.  In
contrast to the neutron scattering, KLV does not loose sensitivity in the protophobic case.\par

The dashed lines show the projected upper bounds of KLV on the product of the couplings
$y_ey_n$, assuming linearity. While Sr/Sr$^+$ will not suffice to probe the couplings relevant for the Be anomaly,
Yb$^+$ has the potential to exclude or support the existence of a new vector with a mass of
$m_X=17\mev$. With a precision of $\mathcal{O}(30\hz)$ Yb$^+$ will become sensitive to the relevant
Be coupling space and with the anticipated precision of $1\hz$ the whole region can be covered.

\section{Conclusions}
\label{sec:concl}

In this work we extended the proposal of Ref.~\cite{Berengut:2017zuo} to constrain New Physics (NP)
by means of isotope shift spectroscopy to enable the inclusion of larger data sets with an arbitrary
number of atomic transitions and isotope pairs. As an application of the King linearity violation
(KLV) observable to bound NP couplings, we evaluated the constraints resulting from existing data
sets of two different atomic systems (Ca$^+$ and Yb).

We compare the existing KLV bounds and near-future projections to present constraints in various
models that can potentially be probed by isotope shifts.
\begin{itemize}
\item \textbf{\emph{B}}$\mathbf{-}$\textbf{\emph{L}:} The $M_{Z'}$-$g_{B-L}$ space is already
  largely constrained by astrophysical and cosmological bounds. Complementary laboratory probes,
  however, are not yet able to confirm those bounds in certain areas of the parameter space. Here
  KLV has the potential to become the strongest laboratory bound for
  $300\ev \lesssim M_{Z'} \lesssim 1\mev$.
\item \textbf{Higgs portal:} While KLV bounds on standard Higgs portals are weaker than existing
  laboratory bounds, KLV can supersede them in the case of an enhanced electron or suppressed
  neutron coupling. For an enhancement (suppression) by a factor of 10, KLV could even set the
  strongest of all bounds in the range $350\kev \lesssim m_\phi \lesssim 1\mev$. Such a scenario can
  be realized e.g.~in the leptonic Higgs portal.\smallskip
\item \textbf{Chameleon:} KLV will be able to set the strongest lower bound $M>500\tev$ on the
  interaction scale of the chameleon with matter.
\item \textbf{Be anomaly:} With the anticipated precision, KLV will fully explore the coupling range
  of a protophobic vector boson with mass $m_X=17\mev$ needed to reproduce the observed anomaly in
  $^8$Be decays.
\end{itemize}

\section*{Acknowledgments}

We thank Cedric Delaunay and Yotam Soreq for useful discussions and for carefully reading our draft.

The work of GP is supported by grants from the BSF, ERC, ISF, Minerva, and the Weizmann- UK Making
Connections Programme.

 \bibliographystyle{utphys}
\bibliography{AtomicNP,AtomicModels,TheBib,IS_light_bib}

\providecommand{\href}[2]{#2}\begingroup\raggedright\begin{thebibliography}{10}

\bibitem{Alexander:2016aln}
J.~Alexander {\em et al.} in {\em {Dark Sectors 2016 Workshop: Community
  Report}}.
\newblock 2016.
\newblock
\href{http://arxiv.org/abs/1608.08632}{{\ttfamily arXiv:1608.08632 [hep-ph]}}.
\newblock

\bibitem{Essig:2009nc}
R.~Essig, P.~Schuster, and N.~Toro, ``{Probing Dark Forces and Light Hidden
  Sectors at Low-Energy e+e- Colliders},''
  \href{http://dx.doi.org/10.1103/PhysRevD.80.015003}{{\em Phys. Rev.}
  {\bfseries D80} (2009) 015003},
\href{http://arxiv.org/abs/0903.3941}{{\ttfamily arXiv:0903.3941 [hep-ph]}}.

\bibitem{Essig:2010xa}
R.~Essig, P.~Schuster, N.~Toro, and B.~Wojtsekhowski, ``{An Electron Fixed
  Target Experiment to Search for a New Vector Boson A' Decaying to e+e-},''
  \href{http://dx.doi.org/10.1007/JHEP02(2011)009}{{\em JHEP} {\bfseries 02}
  (2011) 009},
\href{http://arxiv.org/abs/1001.2557}{{\ttfamily arXiv:1001.2557 [hep-ph]}}.

\bibitem{Batell:2009yf}
B.~Batell, M.~Pospelov, and A.~Ritz, ``{Probing a Secluded U(1) at
  B-factories},'' \href{http://dx.doi.org/10.1103/PhysRevD.79.115008}{{\em
  Phys. Rev.} {\bfseries D79} (2009) 115008},
\href{http://arxiv.org/abs/0903.0363}{{\ttfamily arXiv:0903.0363 [hep-ph]}}.

\bibitem{Batell:2009di}
B.~Batell, M.~Pospelov, and A.~Ritz, ``{Exploring Portals to a Hidden Sector
  Through Fixed Targets},''
  \href{http://dx.doi.org/10.1103/PhysRevD.80.095024}{{\em Phys. Rev.}
  {\bfseries D80} (2009) 095024},
\href{http://arxiv.org/abs/0906.5614}{{\ttfamily arXiv:0906.5614 [hep-ph]}}.

\bibitem{Izaguirre:2014dua}
E.~Izaguirre, G.~Krnjaic, P.~Schuster, and N.~Toro, ``{Physics motivation for a
  pilot dark matter search at Jefferson Laboratory},''
  \href{http://dx.doi.org/10.1103/PhysRevD.90.014052}{{\em Phys. Rev.}
  {\bfseries D90} no.~1, (2014) 014052},
\href{http://arxiv.org/abs/1403.6826}{{\ttfamily arXiv:1403.6826 [hep-ph]}}.

\bibitem{Dobrescu:2014ita}
B.~A. Dobrescu and C.~Frugiuele, ``{GeV-Scale Dark Matter: Production at the
  Main Injector},'' \href{http://dx.doi.org/10.1007/JHEP02(2015)019}{{\em JHEP}
  {\bfseries 02} (2015) 019},
\href{http://arxiv.org/abs/1410.1566}{{\ttfamily arXiv:1410.1566 [hep-ph]}}.

\bibitem{Coloma:2015pih}
P.~Coloma, B.~A. Dobrescu, C.~Frugiuele, and R.~Harnik, ``{Dark matter beams at
  LBNF},'' \href{http://dx.doi.org/10.1007/JHEP04(2016)047}{{\em JHEP}
  {\bfseries 04} (2016) 047},
\href{http://arxiv.org/abs/1512.03852}{{\ttfamily arXiv:1512.03852 [hep-ph]}}.

\bibitem{Frugiuele:2017zvx}
C.~Frugiuele, ``{Probing sub-GeV dark sectors via high energy proton beams at
  LBNF/DUNE and MiniBooNE},''
\href{http://arxiv.org/abs/1701.05464}{{\ttfamily arXiv:1701.05464 [hep-ph]}}.

\bibitem{Flacke:2016szy}
T.~Flacke, C.~Frugiuele, E.~Fuchs, R.~S. Gupta, and G.~Perez, ``{Phenomenology
  of relaxion-Higgs mixing},''
\href{http://arxiv.org/abs/1610.02025}{{\ttfamily arXiv:1610.02025 [hep-ph]}}.

\bibitem{Jaeckel:2010xx}
J.~Jaeckel and S.~Roy, ``{Spectroscopy as a test of Coulomb's law: A Probe of
  the hidden sector},''
  \href{http://dx.doi.org/10.1103/PhysRevD.82.125020}{{\em Phys. Rev.}
  {\bfseries D82} (2010) 125020},
\href{http://arxiv.org/abs/1008.3536}{{\ttfamily arXiv:1008.3536 [hep-ph]}}.

\bibitem{Ficek:2016qwp}
F.~Ficek, D.~F.~J. Kimball, M.~Kozlov, N.~Leefer, S.~Pustelny, and D.~Budker,
  ``{Constraints on exotic spin-dependent interactions between electrons from
  helium fine-structure spectroscopy},''
  \href{http://dx.doi.org/10.1103/PhysRevA.95.032505}{{\em Phys. Rev.}
  {\bfseries A95} no.~3, (2017) 032505},
\href{http://arxiv.org/abs/1608.05779}{{\ttfamily arXiv:1608.05779
  [physics.atom-ph]}}.

\bibitem{Porsev:2009pr}
S.~G. Porsev, K.~Beloy, and A.~Derevianko, ``{Precision determination of
  electroweak coupling from atomic parity violation and implications for
  particle physics},''
  \href{http://dx.doi.org/10.1103/PhysRevLett.102.181601}{{\em Phys. Rev.
  Lett.} {\bfseries 102} (2009) 181601},
\href{http://arxiv.org/abs/0902.0335}{{\ttfamily arXiv:0902.0335 [hep-ph]}}.

\bibitem{Tsigutkin:2009zz}
K.~Tsigutkin, D.~Dounas-Frazer, A.~Family, J.~E. Stalnaker, V.~V. Yashchuk, and
  D.~Budker, ``{Observation of a Large Atomic Parity Violation Effect in
  Ytterbium},''
\href{http://dx.doi.org/10.1103/PhysRevLett.103.071601}{{\em Phys. Rev. Lett.}
  {\bfseries 103} (2009) 071601}.

\bibitem{PhysRevA.81.032114}
K.~Tsigutkin, D.~Dounas-Frazer, A.~Family, J.~E. Stalnaker, V.~V. Yashchuk, and
  D.~Budker, ``Parity violation in atomic ytterbium: Experimental sensitivity
  and systematics,'' \href{http://dx.doi.org/10.1103/PhysRevA.81.032114}{{\em
  Phys. Rev. A} {\bfseries 81} (Mar, 2010) 032114}.

\bibitem{Leefer:2014tga}
N.~Leefer, L.~Bougas, D.~Antypas, and D.~Budker, ``{Towards a new measurement
  of parity violation in dysprosium},'' in {\em {Proceedings of PAVI2014}}.
\newblock 2014.
\newblock
\href{http://arxiv.org/abs/1412.1245}{{\ttfamily arXiv:1412.1245
  [physics.atom-ph]}}.
\newblock

\bibitem{Versolato:2011zza}
O.~O. Versolato {\em et al.}, ``{Atomic parity violation in a single trapped
  radium ion},''
\href{http://dx.doi.org/10.1007/s10751-011-0296-6}{{\em Hyperfine Interact.}
  {\bfseries 199} no.~1-3, (2011) 9--19}.

\bibitem{Dzuba:2012kx}
V.~A. Dzuba, J.~C. Berengut, V.~V. Flambaum, and B.~Roberts, ``{Revisiting
  parity non-conservation in cesium},''
  \href{http://dx.doi.org/10.1103/PhysRevLett.109.203003}{{\em Phys. Rev.
  Lett.} {\bfseries 109} (2012) 203003},
\href{http://arxiv.org/abs/1207.5864}{{\ttfamily arXiv:1207.5864 [hep-ph]}}.

\bibitem{Roberts:2014bka}
B.~M. Roberts, V.~A. Dzuba, and V.~V. Flambaum, ``{Parity and Time-Reversal
  Violation in Atomic Systems},''
  \href{http://dx.doi.org/10.1146/annurev-nucl-102014-022331}{{\em Ann. Rev.
  Nucl. Part. Sci.} {\bfseries 65} (2015) 63--86},
\href{http://arxiv.org/abs/1412.6644}{{\ttfamily arXiv:1412.6644
  [physics.atom-ph]}}.

\bibitem{Jungmann:2014kia}
K.~P. Jungmann, ``{Symmetries and fundamental interactions-selected topics},''
\href{http://dx.doi.org/10.1007/s10751-014-1046-3}{{\em Hyperfine Interact.}
  {\bfseries 227} (2014) 5--16}.

\bibitem{Dzuba:2015xva}
V.~A. Dzuba, V.~V. Flambaum, M.~S. Safronova, S.~G. Porsev, T.~Pruttivarasin,
  M.~A. Hohensee, and H.~H{\"o}ffner, ``{Strongly enhanced effects of Lorentz
  symmetry violation in entangled $\textrm{Yb}^+$ ions},''
\href{http://arxiv.org/abs/1507.06048}{{\ttfamily arXiv:1507.06048
  [physics.atom-ph]}}.

\bibitem{Flambaum:2015rwt}
V.~V. Flambaum, ``{Enhanced effects of the Lorentz invariance and Einstein
  equivalence principle violation in 229Th nuclear transition},''
\href{http://arxiv.org/abs/1511.04848}{{\ttfamily arXiv:1511.04848 [nucl-th]}}.

\bibitem{Godun:2014naa}
R.~Godun, P.~Nisbet-Jones, J.~Jones, S.~King, L.~Johnson, H.~Margolis,
  K.~Szymaniec, S.~Lea, K.~Bongs, and P.~Gill, ``{Frequency Ratio of Two
  Optical Clock Transitions in $^{171}\textrm{Yb}^{+}$ and Constraints on the
  Time Variation of Fundamental Constants},''
\href{http://dx.doi.org/10.1103/PhysRevLett.113.210801}{{\em Phys. Rev. Lett.}
  {\bfseries 113} no.~21, (2014) 210801}.

\bibitem{Rosenband1808}
T.~Rosenband, D.~B. Hume, P.~O. Schmidt, C.~W. Chou, A.~Brusch, L.~Lorini,
  W.~H. Oskay, R.~E. Drullinger, T.~M. Fortier, J.~E. Stalnaker, S.~A. Diddams,
  W.~C. Swann, N.~R. Newbury, W.~M. Itano, D.~J. Wineland, and J.~C. Bergquist,
  ``Frequency Ratio of Al$^+$ and Hg$^+$ Single-Ion Optical Clocks; Metrology
  at the 17th Decimal Place,''
  \href{http://dx.doi.org/10.1126/science.1154622}{{\em Science} {\bfseries
  319} no.~5871, (2008) 1808--1812}.

\bibitem{Delaunay:2016brc}
C.~Delaunay, R.~Ozeri, G.~Perez, and Y.~Soreq, ``{Probing The Atomic Higgs
  Force},''
\href{http://arxiv.org/abs/1601.05087}{{\ttfamily arXiv:1601.05087 [hep-ph]}}.

\bibitem{Berengut:2017zuo}
J.~C. Berengut {\em et al.}, ``{Probing new light force-mediators by isotope
  shift spectroscopy},''
\href{http://arxiv.org/abs/1704.05068}{{\ttfamily arXiv:1704.05068 [hep-ph]}}.

\bibitem{King:63}
W.~H. King, ``Comments on article peculiarities of isotope shift in samarium
  spectrum,'' {\em J. Opt. Soc. Am.} {\bfseries 53} (1963) 638.

\bibitem{King:13}
W.~H. King, {\em Isotope shifts in atomic spectra}.
\newblock Plenum Press, New York, 1984.

\bibitem{Myers2013107}
E.~G. Myers, ``The most precise atomic mass measurements in Penning traps,''
  \href{http://dx.doi.org/10.1016/j.ijms.2013.03.018}{{\em International
  Journal of Mass Spectrometry} {\bfseries 349-350} (2013) 107 -- 122}. 100
  years of Mass Spectrometry.

\bibitem{PhysRevA.94.052511}
M.~Kleinert, M.~E. Gold~Dahl, and S.~Bergeson, ``Measurement of the Yb I
  $^{1}S_{0}\text{\ensuremath{-}}^{1}P_{1}$ transition frequency at 399 nm
  using an optical frequency comb,''
  \href{http://dx.doi.org/10.1103/PhysRevA.94.052511}{{\em Phys. Rev. A}
  {\bfseries 94} (Nov, 2016) 052511}.

\bibitem{PhysRevA.20.239}
D.~L. Clark, M.~E. Cage, D.~A. Lewis, and G.~W. Greenlees, ``Optical isotopic
  shifts and hyperfine splittings for Yb,''
  \href{http://dx.doi.org/10.1103/PhysRevA.20.239}{{\em Phys. Rev. A}
  {\bfseries 20} (Jul, 1979) 239--253}.

\bibitem{PhysRevLett.115.053003}
F.~Gebert, Y.~Wan, F.~Wolf, C.~N. Angstmann, J.~C. Berengut, and P.~O. Schmidt,
  ``Precision Isotope Shift Measurements in Calcium Ions Using Quantum Logic
  Detection Schemes,''
  \href{http://dx.doi.org/10.1103/PhysRevLett.115.053003}{{\em Phys. Rev.
  Lett.} {\bfseries 115} (Jul, 2015) 053003}.

\bibitem{Shi2016}
C.~Shi, F.~Gebert, C.~Gorges, S.~Kaufmann, W.~N{\"o}rtersh{\"a}user, B.~K.
  Sahoo, A.~Surzhykov, V.~A. Yerokhin, J.~C. Berengut, F.~Wolf, J.~C. Heip, and
  P.~O. Schmidt, ``Unexpectedly large difference of the electron density at the
  nucleus in the $4p\, ^2P_{1/2,3/2}$ fine-structure doublet of Ca$^+$,''
  \href{http://dx.doi.org/10.1007/s00340-016-6572-z}{{\em Applied Physics B}
  {\bfseries 123} no.~1, (2016) 2}.

\bibitem{PhysRevA.59.3513}
C.~J. Bowers, D.~Budker, S.~J. Freedman, G.~Gwinner, J.~E. Stalnaker, and
  D.~DeMille, ``Experimental investigation of the
  ${6s}^{2}{}^{1}{S}_{0}\ensuremath{\rightarrow}5d6s{}^{3}{D}_{1,2}$ forbidden
  transitions in atomic ytterbium,''
  \href{http://dx.doi.org/10.1103/PhysRevA.59.3513}{{\em Phys. Rev. A}
  {\bfseries 59} (May, 1999) 3513--3526}.
  \url{https://link.aps.org/doi/10.1103/PhysRevA.59.3513}.

\bibitem{PalmerStacey:1982}
C.~W.~P. Palmer and D.~N. Stacey, ``Theory of anomalous isotope shifts in
  samarium,'' {\em Journal of Physics B: Atomic and Molecular Physics}
  {\bfseries 15} no.~7, (1982) 997.
  \url{http://stacks.iop.org/0022-3700/15/i=7/a=009}.

\bibitem{PhysRev.188.1916}
E.~C. Seltzer, ``{$K$ X-ray} isotope shifts,''
  \href{http://dx.doi.org/10.1103/PhysRev.188.1916}{{\em Phys. Rev.} {\bfseries
  188} (Dec, 1969) 1916--1919}.

\bibitem{SmExp:1981}
J.~A.~R. Griffith, G.~R. Isaak, R.~New, and M.~P. Ralls, ``Anomalies in the
  optical isotope shifts of samarium,'' {\em Journal of Physics B: Atomic and
  Molecular Physics} {\bfseries 14} no.~16, (1981) 2769.
  \url{http://stacks.iop.org/0022-3700/14/i=16/a=007}.

\bibitem{0022-3700-20-15-015}
S.~A. Blundell, P.~E.~G. Baird, C.~W.~P. Palmer, D.~N. Stacey, and G.~K.
  Woodgate, ``A reformulation of the theory of field isotope shift in atoms,''
  {\em Journal of Physics B: Atomic and Molecular Physics} {\bfseries 20}
  no.~15, (1987) 3663. \url{http://stacks.iop.org/0022-3700/20/i=15/a=015}.

\bibitem{PhysRevA.31.2038}
G.~Torbohm, B.~Fricke, and A.~Ros\'en, ``State-dependent volume isotope shifts
  of low-lying states of group-IIa and -IIb elements,''
  \href{http://dx.doi.org/10.1103/PhysRevA.31.2038}{{\em \pr} {\bfseries A31}
  (Apr, 1985) 2038--2053}.

\bibitem{Hogg:2010yz}
D.~W. Hogg, J.~Bovy, and D.~Lang, ``{Data analysis recipes: Fitting a model to
  data},''
\href{http://arxiv.org/abs/1008.4686}{{\ttfamily arXiv:1008.4686
  [astro-ph.IM]}}.

\bibitem{Bordag:2001qi}
M.~Bordag, U.~Mohideen, and V.~M. Mostepanenko, ``{New developments in the
  Casimir effect},''
  \href{http://dx.doi.org/10.1016/S0370-1573(01)00015-1}{{\em Phys. Rept.}
  {\bfseries 353} (2001) 1--205},
\href{http://arxiv.org/abs/quant-ph/0106045}{{\ttfamily arXiv:quant-ph/0106045
  [quant-ph]}}.

\bibitem{bordag2009advances}
M.~Bordag, G.~L. Klimchitskaya, U.~Mohideen, and V.~M. Mostepanenko,
  ``{Advances in the Casimir effect},''
{\em Int. Ser. Monogr. Phys.} {\bfseries 145} (2009) 1--768.

\bibitem{Leeb:1992qf}
H.~Leeb and J.~Schmiedmayer, ``{Constraint on hypothetical light interacting
  bosons from low-energy neutron experiments},''
\href{http://dx.doi.org/10.1103/PhysRevLett.68.1472}{{\em \prl} {\bfseries 68}
  (1992) 1472--1475}.

\bibitem{Nesvizhevsky:2007by}
V.~V. Nesvizhevsky, G.~Pignol, and K.~V. Protasov, ``{Neutron scattering and
  extra short range interactions},''
  \href{http://dx.doi.org/10.1103/PhysRevD.77.034020}{{\em Phys. Rev.}
  {\bfseries D77} (2008) 034020},
\href{http://arxiv.org/abs/0711.2298}{{\ttfamily arXiv:0711.2298 [hep-ph]}}.

\bibitem{Pokotilovski:2006up}
{\relax Yu}.~N. Pokotilovski, ``{Constraints on new interactions from neutron
  scattering experiments},''
  \href{http://dx.doi.org/10.1134/S1063778806060020}{{\em Phys. Atom. Nucl.}
  {\bfseries 69} (2006) 924--931},
\href{http://arxiv.org/abs/hep-ph/0601157}{{\ttfamily arXiv:hep-ph/0601157
  [hep-ph]}}.

\bibitem{PhysRevLett.61.2285}
D.~F. Bartlett and S.~L\"ogl, ``Limits on an Electromagnetic Fifth Force,''
  \href{http://dx.doi.org/10.1103/PhysRevLett.61.2285}{{\em Phys. Rev. Lett.}
  {\bfseries 61} (Nov, 1988) 2285--2287}.

\bibitem{Karshenboim:2010cg}
S.~G. Karshenboim, ``{Precision physics of simple atoms and constraints on a
  light boson with ultraweak coupling},''
  \href{http://dx.doi.org/10.1103/PhysRevLett.104.220406}{{\em Phys. Rev.
  Lett.} {\bfseries 104} (2010) 220406},
\href{http://arxiv.org/abs/1005.4859}{{\ttfamily arXiv:1005.4859 [hep-ph]}}.

\bibitem{Karshenboim:2010ck}
S.~G. Karshenboim, ``{Constraints on a long-range spin-independent interaction
  from precision atomic physics},''
  \href{http://dx.doi.org/10.1103/PhysRevD.82.073003}{{\em Phys. Rev.}
  {\bfseries D82} (2010) 073003},
\href{http://arxiv.org/abs/1005.4872}{{\ttfamily arXiv:1005.4872 [hep-ph]}}.

\bibitem{Harnik:2012ni}
R.~Harnik, J.~Kopp, and P.~A.~N. Machado, ``{Exploring $\nu$ Signals in Dark
  Matter Detectors},''
  \href{http://dx.doi.org/10.1088/1475-7516/2012/07/026}{{\em JCAP} {\bfseries
  1207} (2012) 026},
\href{http://arxiv.org/abs/1202.6073}{{\ttfamily arXiv:1202.6073 [hep-ph]}}.

\bibitem{Andreas:2012mt}
S.~Andreas, C.~Niebuhr, and A.~Ringwald, ``{New Limits on Hidden Photons from
  Past Electron Beam Dumps},''
  \href{http://dx.doi.org/10.1103/PhysRevD.86.095019}{{\em Phys. Rev.}
  {\bfseries D86} (2012) 095019},
\href{http://arxiv.org/abs/1209.6083}{{\ttfamily arXiv:1209.6083 [hep-ph]}}.

\bibitem{Yao:2006px}
{\bfseries Particle Data Group} Collaboration, W.~M. Yao {\em et al.},
  ``{Review of Particle Physics},''
\href{http://dx.doi.org/10.1088/0954-3899/33/1/001}{{\em J. Phys.} {\bfseries
  G33} (2006) 1--1232}.

\bibitem{Raffelt:2012sp}
G.~Raffelt, ``{Limits on a CP-violating scalar axion-nucleon interaction},''
  \href{http://dx.doi.org/10.1103/PhysRevD.86.015001}{{\em Phys. Rev.}
  {\bfseries D86} (2012) 015001},
\href{http://arxiv.org/abs/1205.1776}{{\ttfamily arXiv:1205.1776 [hep-ph]}}.

\bibitem{Blum:2016afe}
K.~Blum and D.~Kushnir, ``{Neutrino Signal of Collapse-induced Thermonuclear
  Supernovae: the Case for Prompt Black Hole Formation in SN1987A},''
  \href{http://dx.doi.org/10.3847/0004-637X/828/1/31}{{\em Astrophys. J.}
  {\bfseries 828} no.~1, (2016) 31},
\href{http://arxiv.org/abs/1601.03422}{{\ttfamily arXiv:1601.03422
  [astro-ph.HE]}}.

\bibitem{Grifols:1986fc}
J.~A. Grifols and E.~Masso, ``{Constraints on Finite Range Baryonic and
  Leptonic Forces From Stellar Evolution},''
\href{http://dx.doi.org/10.1016/0370-2693(86)90509-5}{{\em Phys. Lett.}
  {\bfseries B173} (1986) 237--240}.

\bibitem{Grifols:1988fv}
J.~A. Grifols, E.~Masso, and S.~Peris, ``{Energy Loss From the Sun and {RED}
  Giants: Bounds on Short Range Baryonic and Leptonic Forces},''
\href{http://dx.doi.org/10.1142/S0217732389000381}{{\em Mod. Phys. Lett.}
  {\bfseries A4} (1989) 311}.

\bibitem{Hardy:2016kme}
E.~Hardy and R.~Lasenby, ``{Stellar cooling bounds on new light particles:
  plasma mixing effects},''
  \href{http://dx.doi.org/10.1007/JHEP02(2017)033}{{\em JHEP} {\bfseries 02}
  (2017) 033},
\href{http://arxiv.org/abs/1611.05852}{{\ttfamily arXiv:1611.05852 [hep-ph]}}.

\bibitem{Redondo:2013lna}
J.~Redondo and G.~Raffelt, ``{Solar constraints on hidden photons
  re-visited},'' \href{http://dx.doi.org/10.1088/1475-7516/2013/08/034}{{\em
  JCAP} {\bfseries 1308} (2013) 034},
\href{http://arxiv.org/abs/1305.2920}{{\ttfamily arXiv:1305.2920 [hep-ph]}}.

\bibitem{Ahlgren:2013wba}
B.~Ahlgren, T.~Ohlsson, and S.~Zhou, ``{Comment on ``Is Dark Matter with
  Long-Range Interactions a Solution to All Small-Scale Problems of $\Lambda$
  Cold Dark Matter Cosmology''},''
  \href{http://dx.doi.org/10.1103/PhysRevLett.111.199001}{{\em Phys. Rev.
  Lett.} {\bfseries 111} no.~19, (2013) 199001},
\href{http://arxiv.org/abs/1309.0991}{{\ttfamily arXiv:1309.0991 [hep-ph]}}.

\bibitem{Heeck:2014zfa}
J.~Heeck, ``{Unbroken B – L symmetry},''
  \href{http://dx.doi.org/10.1016/j.physletb.2014.10.067}{{\em Phys. Lett.}
  {\bfseries B739} (2014) 256--262},
\href{http://arxiv.org/abs/1408.6845}{{\ttfamily arXiv:1408.6845 [hep-ph]}}.

\bibitem{Jaeckel:2010ni}
J.~Jaeckel and A.~Ringwald, ``{The Low-Energy Frontier of Particle Physics},''
  \href{http://dx.doi.org/10.1146/annurev.nucl.012809.104433}{{\em Ann. Rev.
  Nucl. Part. Sci.} {\bfseries 60} (2010) 405--437},
\href{http://arxiv.org/abs/1002.0329}{{\ttfamily arXiv:1002.0329 [hep-ph]}}.

\bibitem{Barbieri:1975xy}
R.~Barbieri and T.~E.~O. Ericson, ``{Evidence Against the Existence of a Low
  Mass Scalar Boson from Neutron-Nucleus Scattering},''
\href{http://dx.doi.org/10.1016/0370-2693(75)90073-8}{{\em Phys. Lett.}
  {\bfseries 57B} (1975) 270--272}.

\bibitem{Wissmann:1998ta}
F.~Wissmann, M.~Schumacher, and M.~I. Levchuk, ``{On approaches to measure the
  electromagnetic polarizabilities of the neutron},''
{\em Eur. Phys. J.} {\bfseries A1} (1998) 193--200.

\bibitem{Antoniadis:2011zza}
I.~Antoniadis {\em et al.}, ``{Short-range fundamental forces},''
\href{http://dx.doi.org/10.1016/j.crhy.2011.05.004}{{\em Comptes Rendus
  Physique} {\bfseries 12} (2011) 755--778}.

\bibitem{TuckerSmith:2010ra}
D.~Tucker-Smith and I.~Yavin, ``{Muonic hydrogen and MeV forces},''
  \href{http://dx.doi.org/10.1103/PhysRevD.83.101702}{{\em Phys. Rev.}
  {\bfseries D83} (2011) 101702},
\href{http://arxiv.org/abs/1011.4922}{{\ttfamily arXiv:1011.4922 [hep-ph]}}.

\bibitem{Bellini:2011rx}
G.~Bellini {\em et al.}, ``{Precision measurement of the 7Be solar neutrino
  interaction rate in Borexino},''
  \href{http://dx.doi.org/10.1103/PhysRevLett.107.141302}{{\em Phys. Rev.
  Lett.} {\bfseries 107} (2011) 141302},
\href{http://arxiv.org/abs/1104.1816}{{\ttfamily arXiv:1104.1816 [hep-ex]}}.

\bibitem{Beda:2009kx}
A.~G. Beda, E.~V. Demidova, A.~S. Starostin, V.~B. Brudanin, V.~G. Egorov,
  D.~V. Medvedev, M.~V. Shirchenko, and T.~Vylov, ``{GEMMA experiment: Three
  years of the search for the neutrino magnetic moment},''
  \href{http://dx.doi.org/10.1134/S1547477110060063}{{\em Phys. Part. Nucl.
  Lett.} {\bfseries 7} (2010) 406--409},
\href{http://arxiv.org/abs/0906.1926}{{\ttfamily arXiv:0906.1926 [hep-ex]}}.

\bibitem{PhysRevLett.100.120801}
D.~Hanneke, S.~Fogwell, and G.~Gabrielse, ``New Measurement of the Electron
  Magnetic Moment and the Fine Structure Constant,''
  \href{http://dx.doi.org/10.1103/PhysRevLett.100.120801}{{\em \prl} {\bfseries
  100} (Mar, 2008) 120801}.

\bibitem{Dobrescu:2014fca}
B.~A. Dobrescu and C.~Frugiuele, ``{Hidden GeV-scale interactions of quarks},''
  \href{http://dx.doi.org/10.1103/PhysRevLett.113.061801}{{\em Phys. Rev.
  Lett.} {\bfseries 113} (2014) 061801},
\href{http://arxiv.org/abs/1404.3947}{{\ttfamily arXiv:1404.3947 [hep-ph]}}.

\bibitem{Kozaczuk:2016nma}
J.~Kozaczuk, D.~E. Morrissey, and S.~R. Stroberg, ``{Light Axial Vectors,
  Nuclear Transitions, and the $^8$Be Anomaly},''
\href{http://arxiv.org/abs/1612.01525}{{\ttfamily arXiv:1612.01525 [hep-ph]}}.

\bibitem{Kahn:2016vjr}
Y.~Kahn, G.~Krnjaic, S.~Mishra-Sharma, and T.~M.~P. Tait, ``{Light Weakly
  Coupled Axial Forces: Models, Constraints, and Projections},''
\href{http://arxiv.org/abs/1609.09072}{{\ttfamily arXiv:1609.09072 [hep-ph]}}.

\bibitem{Bouchiat:2004sp}
C.~Bouchiat and P.~Fayet, ``{Constraints on the parity-violating couplings of a
  new gauge boson},''
  \href{http://dx.doi.org/10.1016/j.physletb.2004.12.065}{{\em Phys. Lett.}
  {\bfseries B608} (2005) 87--94},
\href{http://arxiv.org/abs/hep-ph/0410260}{{\ttfamily arXiv:hep-ph/0410260
  [hep-ph]}}.

\bibitem{Davoudiasl:2012ag}
H.~Davoudiasl, H.-S. Lee, and W.~J. Marciano, ``{'Dark' Z implications for
  Parity Violation, Rare Meson Decays, and Higgs Physics},''
  \href{http://dx.doi.org/10.1103/PhysRevD.85.115019}{{\em Phys. Rev.}
  {\bfseries D85} (2012) 115019},
\href{http://arxiv.org/abs/1203.2947}{{\ttfamily arXiv:1203.2947 [hep-ph]}}.

\bibitem{Patt:2006fw}
B.~Patt and F.~Wilczek, ``{Higgs-field portal into hidden sectors},''
\href{http://arxiv.org/abs/hep-ph/0605188}{{\ttfamily arXiv:hep-ph/0605188
  [hep-ph]}}.

\bibitem{OConnell:2006rsp}
D.~O'Connell, M.~J. Ramsey-Musolf, and M.~B. Wise, ``{Minimal Extension of the
  Standard Model Scalar Sector},''
  \href{http://dx.doi.org/10.1103/PhysRevD.75.037701}{{\em Phys. Rev.}
  {\bfseries D75} (2007) 037701},
\href{http://arxiv.org/abs/hep-ph/0611014}{{\ttfamily arXiv:hep-ph/0611014
  [hep-ph]}}.

\bibitem{Choi:2016luu}
K.~Choi and S.~H. Im, ``{Constraints on Relaxion Windows},''
  \href{http://dx.doi.org/10.1007/JHEP12(2016)093}{{\em JHEP} {\bfseries 12}
  (2016) 093},
\href{http://arxiv.org/abs/1610.00680}{{\ttfamily arXiv:1610.00680 [hep-ph]}}.

\bibitem{Belanger:2008sj}
G.~Belanger, F.~Boudjema, A.~Pukhov, and A.~Semenov, ``{Dark matter direct
  detection rate in a generic model with micrOMEGAs 2.2},''
  \href{http://dx.doi.org/10.1016/j.cpc.2008.11.019}{{\em Comput. Phys.
  Commun.} {\bfseries 180} (2009) 747--767},
\href{http://arxiv.org/abs/0803.2360}{{\ttfamily arXiv:0803.2360 [hep-ph]}}.

\bibitem{Belanger:2013oya}
G.~Belanger, F.~Boudjema, A.~Pukhov, and A.~Semenov, ``{micrOMEGAs 3: A program
  for calculating dark matter observables},''
  \href{http://dx.doi.org/10.1016/j.cpc.2013.10.016}{{\em Comput. Phys.
  Commun.} {\bfseries 185} (2014) 960--985},
\href{http://arxiv.org/abs/1305.0237}{{\ttfamily arXiv:1305.0237 [hep-ph]}}.

\bibitem{Shifman:1978zn}
M.~A. Shifman, A.~I. Vainshtein, and V.~I. Zakharov, ``{Remarks on Higgs Boson
  Interactions with Nucleons},''
\href{http://dx.doi.org/10.1016/0370-2693(78)90481-1}{{\em Phys. Lett.}
  {\bfseries B78} (1978) 443}.

\bibitem{Batell:2016ove}
B.~Batell, N.~Lange, D.~McKeen, M.~Pospelov, and A.~Ritz, ``{Muon anomalous
  magnetic moment through the leptonic Higgs portal},''
  \href{http://dx.doi.org/10.1103/PhysRevD.95.075003}{{\em Phys. Rev.}
  {\bfseries D95} no.~7, (2017) 075003},
\href{http://arxiv.org/abs/1606.04943}{{\ttfamily arXiv:1606.04943 [hep-ph]}}.

\bibitem{Khoury:2003aq}
J.~Khoury and A.~Weltman, ``{Chameleon fields: Awaiting surprises for tests of
  gravity in space},''
  \href{http://dx.doi.org/10.1103/PhysRevLett.93.171104}{{\em Phys. Rev. Lett.}
  {\bfseries 93} (2004) 171104},
\href{http://arxiv.org/abs/astro-ph/0309300}{{\ttfamily arXiv:astro-ph/0309300
  [astro-ph]}}.

\bibitem{Khoury:2003rn}
J.~Khoury and A.~Weltman, ``{Chameleon cosmology},''
  \href{http://dx.doi.org/10.1103/PhysRevD.69.044026}{{\em Phys. Rev.}
  {\bfseries D69} (2004) 044026},
\href{http://arxiv.org/abs/astro-ph/0309411}{{\ttfamily arXiv:astro-ph/0309411
  [astro-ph]}}.

\bibitem{Brax:2010jk}
P.~Brax and C.~Burrage, ``{Chameleon Induced Atomic Afterglow},''
  \href{http://dx.doi.org/10.1103/PhysRevD.82.095014}{{\em Phys. Rev.}
  {\bfseries D82} (2010) 095014},
\href{http://arxiv.org/abs/1009.1065}{{\ttfamily arXiv:1009.1065 [hep-ph]}}.

\bibitem{Brax:2010gp}
P.~Brax and C.~Burrage, ``{Atomic Precision Tests and Light Scalar
  Couplings},'' \href{http://dx.doi.org/10.1103/PhysRevD.83.035020}{{\em Phys.
  Rev.} {\bfseries D83} (2011) 035020},
\href{http://arxiv.org/abs/1010.5108}{{\ttfamily arXiv:1010.5108 [hep-ph]}}.

\bibitem{Burrage:2014oza}
C.~Burrage, E.~J. Copeland, and E.~A. Hinds, ``{Probing Dark Energy with Atom
  Interferometry},''
  \href{http://dx.doi.org/10.1088/1475-7516/2015/03/042}{{\em JCAP} {\bfseries
  1503} no.~03, (2015) 042},
\href{http://arxiv.org/abs/1408.1409}{{\ttfamily arXiv:1408.1409
  [astro-ph.CO]}}.

\bibitem{Schwob:1999zz}
C.~Schwob, L.~Jozefowski, B.~de~Beauvoir, L.~Hilico, F.~Nez, L.~Julien,
  F.~Biraben, O.~Acef, J.~J. Zondy, and A.~Clairon, ``{Optical Frequency
  Measurement of the S-2- D-12 Transitions in Hydrogen and Deuterium: Rydberg
  Constant and Lamb Shift Determinations},''
\href{http://dx.doi.org/10.1103/PhysRevLett.82.4960}{{\em Phys. Rev. Lett.}
  {\bfseries 82} (1999) 4960--4963}.

\bibitem{Simon:1980hu}
G.~G. Simon, C.~Schmitt, F.~Borkowski, and V.~H. Walther, ``{Absolute electron
  Proton Cross-Sections at Low Momentum Transfer Measured with a High Pressure
  Gas Target System},''
\href{http://dx.doi.org/10.1016/0375-9474(80)90104-9}{{\em Nucl. Phys.}
  {\bfseries A333} (1980) 381--391}.

\bibitem{Burrage:2016bwy}
C.~Burrage and J.~Sakstein, ``{A Compendium of Chameleon Constraints},''
  \href{http://dx.doi.org/10.1088/1475-7516/2016/11/045}{{\em JCAP} {\bfseries
  1611} no.~11, (2016) 045},
\href{http://arxiv.org/abs/1609.01192}{{\ttfamily arXiv:1609.01192
  [astro-ph.CO]}}.

\bibitem{Krasznahorkay:2015iga}
A.~J. Krasznahorkay {\em et al.}, ``{Observation of Anomalous Internal Pair
  Creation in Be8 : A Possible Indication of a Light, Neutral Boson},''
  \href{http://dx.doi.org/10.1103/PhysRevLett.116.042501}{{\em Phys. Rev.
  Lett.} {\bfseries 116} no.~4, (2016) 042501},
\href{http://arxiv.org/abs/1504.01527}{{\ttfamily arXiv:1504.01527 [nucl-ex]}}.

\bibitem{Feng:2016jff}
J.~L. Feng, B.~Fornal, I.~Galon, S.~Gardner, J.~Smolinsky, T.~M.~P. Tait, and
  P.~Tanedo, ``{Protophobic Fifth-Force Interpretation of the Observed Anomaly
  in $^8$Be Nuclear Transitions},''
  \href{http://dx.doi.org/10.1103/PhysRevLett.117.071803}{{\em Phys. Rev.
  Lett.} {\bfseries 117} no.~7, (2016) 071803},
\href{http://arxiv.org/abs/1604.07411}{{\ttfamily arXiv:1604.07411 [hep-ph]}}.

\bibitem{Feng:2016ysn}
J.~L. Feng, B.~Fornal, I.~Galon, S.~Gardner, J.~Smolinsky, T.~M.~P. Tait, and
  P.~Tanedo, ``{Particle physics models for the 17 MeV anomaly in beryllium
  nuclear decays},'' \href{http://dx.doi.org/10.1103/PhysRevD.95.035017}{{\em
  Phys. Rev.} {\bfseries D95} no.~3, (2017) 035017},
\href{http://arxiv.org/abs/1608.03591}{{\ttfamily arXiv:1608.03591 [hep-ph]}}.

\bibitem{Anastasi:2015qla}
A.~Anastasi {\em et al.}, ``{Limit on the production of a low-mass vector boson
  in $\mathrm{e}^{+}\mathrm{e}^{-} \to \mathrm{U}\gamma$, $\mathrm{U} \to
  \mathrm{e}^{+}\mathrm{e}^{-}$ with the KLOE experiment},''
  \href{http://dx.doi.org/10.1016/j.physletb.2015.10.003}{{\em Phys. Lett.}
  {\bfseries B750} (2015) 633--637},
\href{http://arxiv.org/abs/1509.00740}{{\ttfamily arXiv:1509.00740 [hep-ex]}}.

\bibitem{Riordan:1987aw}
E.~M. Riordan {\em et al.}, ``{A Search for Short Lived Axions in an Electron
  Beam Dump Experiment},''
\href{http://dx.doi.org/10.1103/PhysRevLett.59.755}{{\em Phys. Rev. Lett.}
  {\bfseries 59} (1987) 755}.

\end{thebibliography}\endgroup
\markboth{}{}

\end{document}